# Photonic Terahertz Phased Array

Li Niu[1,#], Xi Feng[1,#], Xueqian Zhang[1,*], Yongchang Lu[1], Qingwei Wang[1], Quan Xu[1], Xieyu Chen[1], Jiajun Ma[1], Haidi Qiu[1], Wei E.I. Sha[2], Shuang Zhang[3], Andrea Alù[4,5,*], Weili Zhang[6,*], Jiaguang Han[1,7,*]

[1]Center for Terahertz Waves and College of Precision Instrument and Optoelectronics Engineering, and the Key Laboratory of Optoelectronics Information and Technology, Tianjin University, Tianjin 300072, P. R. China
[2]Key Laboratory of Micro-nano Electronic Devices and Smart Systems of Zhejiang Province, College of Information Science & Electronic Engineering, Zhejiang University, Hangzhou 310027, China;
[3]Department of Electrical & Electronic Engineering, University of Hong Kong; 999077, Hong Kong, China
[4]Photonics Initiative, Advanced Science Research Center, City University of New York, New York, 10031, USA
[5]Physics Program, Graduate Center, City University of New York, New York, 10016, USA.
[6]School of Electrical and Computer Engineering, Oklahoma State University, Stillwater, Oklahoma 74078, USA
[7]Guangxi Key Laboratory of Optoelectronic Information Processing, School of Optoelectronic Engineering, Guilin University of Electronic Technology, Guilin 541004, China
[#]These authors contributed equally to this work
[*]Corresponding authors. Email: alearn1988@tju.edu.cn, aalu@gc.cuny.edu, weili.zhang@okstate.edu, jiaghan@tju.edu.cn

## Abstract

Phased arrays are crucial in various technologies, such as radar and wireless communications, due to their ability to precisely control and steer electromagnetic waves. This precise control improves signal processing and enhances imaging performance. However, extending phased arrays to the terahertz (THz) frequency range has proven challenging, especially for high-frequency operation, broadband performance, two-dimensional (2D) phase control with large antenna arrays, and strong phase modulation. Here, we introduce a photonic platform to realize a THz phased array that bypasses the above challenges. Our method employs 2D phase coding with 2-bit across a broad THz frequency range from 0.8 to 1.4 THz. The core of our design is a pixelated nonlinear Pancharatnam-Berry metasurface driven by a spatially modulated femtosecond laser, allowing precise phase control of THz signals. We showcase the effectiveness of our method through four proof-of-concept applications: single beamforming, dual beamforming, imaging and vortex beam generation. The realized photonic platform provides a promising pathway for developing broadband phased arrays in the THz regime.

**Introduction**

A phased array is a system featuring multiple antennas that can dynamically steer the direction of the radiated electromagnetic beam. This is typically accomplished by introducing time variations or phase delays in the signal paths of each antenna, effectively compensating for differences in the signal paths through free space.[1,2] The versatility and efficiency of phased arrays make them a crucial component to a wide range of advanced technological systems and applications ranging from radar, communication, and astronomy.[3,4]

Research on phased arrays can be dated back to the initial development of radio frequency technologies. Over time, advancements in microwave and millimeter-wave technologies have significantly influenced their evolution.[5,6] Recently, emerging communication technologies operating at millimeter-wave and terahertz (THz) frequencies have attracted widespread attention due to superior bandwidth, directivity and resolution compared to microwaves.[7] In order to enhance the performance of wireless communication systems at these higher frequencies, new phased array architectures with superior capabilities are anticipated to play a critical role.

Traditional electronic phased array systems, typically relying on coupled oscillators and phase shifters for phase adjustment, are preferred at millimeter-wave and lower THz frequencies.[8,9] However, as the frequencies increase, current electronic devices face inherent delays and restricted bandwidths, hindering their capabilities.[10,11] Transmission losses and intermodulation interference further degrade their performance at high frequencies. In addition, these systems encounter problems with low cutoff frequencies, transistor attenuation, and high manufacturing costs, with

current systems predominantly operating below 400 GHz.[12,13]

Photonic approaches present a promising solution to bypass the challenges faced by electronic phase arrays. Photonic THz sources, including nonlinear crystals,[14] spintronic materials,[15,16] topological materials,[17,18] and photoconductive antennas[19] can generate broadband THz waves, reaching frequencies of several THz. However, a significant challenge in implementing a phased array is the active control of the phase of the generated THz signals. One potential solution involves integrating a THz source with a programmable linear metasurface. This approach, relying on the use of active phase components such as vanadium oxide,[20] graphene,[21] high-electron-mobility transistors,[22] and liquid crystals,[23,24] has significantly advanced the development of terahertz manipulation technologies. However, it also faces some issues, such as complex control systems, complicated power supply arrangements, and unwanted interference. Furthermore, the reliance on resonant interactions often restricts the operational bandwidth and limits the phase resolution. Additionally, separating the THz source from the phase control device also introduces additional insertion losses, reducing the system's overall integration.[13]

Nonlinear metasurfaces have emerged as a promising solution for precise control of THz wavefronts by utilizing the nonlinear Pancharatnam-Berry (PB) phase.[25-27] They have enabled advanced control of terahertz waves such as holography,[28] polarization,[29] and toroidal beam generation.[30] However, implementing THz phased arrays requires additional mechanisms for dynamic phase control. In this article, we introduce a novel approach for creating a programmable broadband photonic THz phased array (PTPA) utilizing nonlinear metasurfaces. Our method employs a Digital Micromirror Device (DMD) to selectively activate specific phase control elements within the nonlinear metasurface, thereby allowing for the customized generation of THz wavefronts. This method significantly simplifies both the design and control processes for enabling real-time generation of diverse functionalities. Through proof-of-concept experiments, we demonstrate broadband THz generation, precise single and dual beamforming, imaging capabilities, and vortex beams creation. These results highlight the potential of this approach for a wide range of THz applications.

## Results

Our PTPA design strategy is illustrated in Fig. 1. It involves a two-dimensional (2D) pixelated structure composed of split-ring resonator (SRR) arrays, with elements oriented at 0°, 90°, 180°, and 270°. According to nonlinear PB phase, when pumped by a given CP infrared beam, the generated left-handed circularly polarized (LCP) THz waves acquire phases of 0°, 90°, 180°, and 270°, while right-handed circularly polarized (RCP) THz waves acquire phases of 0°, −90°, −180°, and −270°, depending on the orientation of the excited SRR arrays. This arrangement facilitates a 2-bit PTPA, offering four phase codes: "00", "01", "10", and "11". The utilization of a DMD as a high-speed spatial light modulator enables flexible encoding of the pattern of the pump beam—a near-infrared, circularly polarized femtosecond laser beam—to selectively excite the desired SRR elements. Consequently, advanced programmable broadband THz wavefront control is achieved.

Figure 2a illustrates the basic building block of the PTPA, which consists of a gold SRR fabricated on an indium-tin-oxide (ITO)-coated glass substrate. Each basic unit is an array of 130 × 130 SRRs with a period of $P_1$ = 382 nm, forming a sub-element (with a size of 50 μm × 50 μm, exhibiting a magnetic resonance at 1275 nm as depicted in Fig. S1). These sub-elements are then arranged in a 4 × 4 grid, and the SRRs are oriented at 0°, 90°, 180°, and 270°, constituting a PTPA element (with a size of 200 μm × 200 μm), as depicted in Fig. 2b. The final fabricated PTPA sample, which includes 10 × 5 these elements, covers a total area of 2 mm × 1 mm. A partial scanning electron microscope image of the sample is illustrated in Fig. 2c.

The dynamic 2-bit phase coding mechanism of the PTPA element is depicted in Fig. 2d. It relies on programming a binary coding pattern into the DMD. This pattern modulates the pump beam to selectively excite one of the four sub-elements within the PTPA, leading to the generation of THz waves with a consistent nonlinear PB phase. Although the excited sub-elements are positioned differently, which induces phase differences for emitted THz waves towards oblique directions. Given that the sub-elements are subwavelength and their spatial offsets are small compared to the element size, this phase crosstalk is not considered in Fig. 2d for simplicity. This would not affect the fundamental controlling picture, since the main phase contribution is clearly arisen from the nonlinear PB phase and our following results also support this.

To better illustrate the DMD coding process, Fig. 2e provides an enlarged view of the micromirror arrangement used to control two sub-elements in the on-state and off-state. Each micromirror has a diagonal length of 10.8 μm. Due to the diamond shape of the micromirrors, the coding area appears square but with slight variations along the edges. During operation, each micromirror alternates between an angle of −12° (on-state) and 12° (off-state), as depicted in Fig 2f. Only the laser beam that is reflected in the desired direction is collected by the PTPA.

**Dynamic phase control of the PTPA element**

To validate the proposed PTPA, we established a DMD-integrated THz time-domain spectroscopy system, as detailed in Supplementary Note 1. This setup enables the projection of various pump beam patterns onto the PTPA, exciting different phase states of the elements and thus creating diverse phase distributions for the generated THz waves. This allows for the direct generation of desired THz wavefronts. Here, all the measurements are performed using an LCP pump. Figure 2g illustrates the measured normalized broadband active phase control response for the generated LCP and RCP THz waves at 1275-nm pump wavelength. In this case, all the elements are sequentially excited using the same patterns shown in Fig. 2d. The phase responses are stably controlled over a broadband frequency range from 0.8 to 1.4 THz. As expected, the phase responses for LCP and RCP THz waves are inverted, with the "00" state serving as the reference. Specifically, the phase of LCP THz waves increases from 0° to 270° in 90° increments, while the phase of RCP THz waves decreases from 0° to −270° in −90° increments. These results confirm the 2-bit broadband phase coding capability of our PTPA element design. The corresponding time-domain THz pulses used to derive these results are shown in Fig. S3.

**Programmable THz single beamforming**

We first explore the capability of our PTPA as a THz single beamforming device. The 2D arrangement of the elements in principle allows 2D scanning of the generated THz beam by designing appropriate phase gradients. For simplicity, we focus on a one-dimensional (1D) single beamforming along the *x*-direction as a proof-of-concept demonstration, as illustrated in Fig. 3a. The deflection angle of the beam is controlled by designing an interfacial phase gradient d$\Phi$/d*x*, which corresponds to the following phase distribution:

$$\varphi(x) = \frac{d\Phi}{dx} x. \tag{1}$$

By discretizing the phase into four levels and the *x*-coordinate according to the resolution of the elements, we can encode the phase using the method in Fig. 2d. Figures 3b and d show two exemplary DMD coding patterns applied to the pump beam, referred as case 1 and case 2, respectively. In case 1, we set d$\Phi_1$/d*x* = $2\pi/\Lambda_1$ with $\Lambda_1$ = 1.6 mm, while in case 2, d$\Phi_2$/d*x* = $2\pi/\Lambda_2$ with $\Lambda_2$ = 0.8 mm. The periodically tilt observed in the on-states results from the specific arrangement of sub-elements. Their corresponding phase coding distributions are depicted in Figs. 3c and e, which clearly show linear phase variations. For these two cases, the PTPA exhibits 1.25 and 2.5 phase periods, respectively.

We first carry out numerical calculations to investigate the corresponding 1D far-field radiations using[31]:

$$E(\alpha) = \sum_{m=1}^{N_x} \sum_{n=1}^{N_y} I(m,n) \exp\{i[\varphi(m,n) + kP_2(m-1/2)\sin(\alpha)]\}, \tag{2}$$

where $N_x$ and $N_y$ denote the number of subelements in columns and rows. $I(m, n)$ = 0 or 1 represents whether the subelement at position (*m*, *n*) is excited, $\varphi(m, n)$ is the phase of the subelement, $k = 2\pi/\lambda$ is the wave number with $\lambda$ being the wavelength, and $\alpha$ is the deflection angle of the generated THz beam.

Figure 3f illustrates the calculated angle-resolved LCP THz intensity distributions for the two cases at 1.0, 1.2 and 1.4 THz, respectively. To align with the experimental results, these intensity distributions have been normalized according to the intensity ratios of the THz waves generated by the uniform SRR array at different frequencies. The results reveal clear single THz beamforming. Due to the different magnitudes of the phase gradients, the deflection angles vary accordingly. Figure 3g presents the corresponding experimental results, which agree well with the calculations. The range of deflection angle $\alpha$, from $-28°$ to $18°$, is constrained by the numerical aperture of the THz polarizer and parabolic mirror used for collecting the THz radiation. Both the calculated and measured deflection angles align with the theoretical predictions from the generalized Snell's law,[32] $\sin(\alpha) = 1/k \times d\Phi/dx = \lambda/\Lambda$, as indicated by the dashed lines in Figs. 3f and g. Taking 1.0 THz as an example, the deflection angles are $-10.8°$ and $-22°$ for case 1 and case 2, respectively.

In a phased array, beamwidth $\Omega$ and directivity $D$ are two critical parameters. For our 1D demonstration, the theoretical values are calculated using $\Omega_{theo.} = \lambda/L\cos(\alpha)$ and $D_{theo.} = 4\pi/\Omega_{theo.}$, where $L$ is the length of the PTPA.[1] Substituting the relevant values yields theoretical beamwidths $\Omega_{theo.} = 8.7°$ and $9.3°$, and directivity $D_{theo.} = 82.3$ and $77.7$ for case 1 and case 2, respectively. In practice, $\Omega$ is defined by the full width at half maximum (FWHM) of the radiated beam. Based on the results shown in Figs. 3f and g, the beamwidth can be estimated as $\Omega_{cal.} = 7.8°$ and $8.3°$, $D_{cal.} = 92.4$ and $86.6$, while experimental values are $\Omega_{exp.} = 15.6°$ and $13.8°$, $D_{exp.} = 46.2$ and $56.1$, respectively. The calculated results are close to the theoretical predictions, implying the effectiveness of our approach in principle. The deviations observed in the experimental results, such as the increased beamwidths and decreased directivities, can be attributed to the limitations of the signal-to-noise ratio and frequency resolution of our current experimental setup. Despite these differences, the values can still serve as useful references without altering the underlying physics. Theoretically, using the coding method is Fig. 2d, the deflection angle range of our PTPA at 1.0 THz can reach $[-48.6°, 48.6°]$. This range can be further expanded by optimizing the coding style.

Additionally, we investigate the frequency bandwidth $\Delta f$ of the two cases at specific deflection angles, which are important for practical applications as they determine the maximum available modulation speed. Figures 3h and i depict the calculated and measured intensity spectra of the generated LCP THz waves at $-25°$, $-20°$ and $-15°$, respectively. The main features and trends of these spectra are in good agreement with each other. The dashed lines indicate the theoretical frequencies at corresponding deflection angles. The calculated and experimentally measured frequency bandwidths are $\Delta f_{cal.}$ ($\Delta f_{exp.}$) = 0.27 (0.39), 0.35 (0.40) and 0.40 (0.43) THz for case 1, while 0.26 (0.41), 0.31 (0.63) and 0.43 (0.8) THz for case 2, respectively. For the generated RCP THz waves, the above results exhibit deflections towards the opposite angles due to their phase conjugate relation with the LCP THz waves, as shown in Fig. S4.

**Programmable THz dual beamforming**

Multiple beamforming enables the generation of multiple THz beams, each directed at different users, facilitating simultaneous services or tasks.[33] This is particularly valuable for enhancing the capacity and efficiency of THz wireless communication and radar applications. Next, we demonstrate the dual beamforming functionality of our PTPA, as shown in Fig. 4a, where we can independently control the deflection angles of the two beams. We use LCP THz beam generation as an example, noting that the RCP counterparts are mirror images of the LCP beams with respect to the normal direction. The crucial aspect lies in establishing appropriate phase distributions within the PTPA, achieved through a holographic method that involves superimposing two obliquely propagating beams as:

$$\varphi(x) = arg\left\{ A_1 \exp\left[i\left(\frac{d\Phi_1}{dx}x\right)\right] + A_2 \exp\left[i\left(\frac{d\Phi_2}{dx}x\right)\right]\right\}, \qquad (3)$$

where $A_1$, $A_2$ represent the initial amplitudes of the two desired beams, while $d\Phi_1/dx$, $d\Phi_2/dx$ represent their interfacial phase gradients, respectively. Since amplitude control is excluded here, $A_1$ and $A_2$ cannot represent the real amplitude relation between the two output THz beams. To ensure two similar amplitudes, we optimize the far field radiation pattern using Eq. (2), selecting $A_1 = 0.97$ and $A_2 = 1$. Figures 4b and d show two exemplary DMD coding patterns for the pump beam, denoted as case 3 with $d\Phi_1/dx = 0$ mm and $d\Phi_2/dx = -2\pi/1.6$, and case 4 with $d\Phi_1/dx = 2\pi/0.8$ mm and $d\Phi_2/dx = -2\pi/1.6$ mm, respectively. The corresponding equivalent phase coding distributions are illustrated in Figs. 4c and e. The spatially non-uniform on-states and phase distributions result from the interference between the two desired beams.

Figures 4f and h show the calculated angle-resolved LCP THz intensity distributions at 1.0, 1.2, and 1.4 THz, respectively. Clear dual THz beams with nearly the same amplitude are observed. Figures 4g and i present the measured results. The corresponding deflection angles of them are all consistent with the theoretical calculations, as indicated by the dashed lines. We also compared the theoretical, calculated, and experimental beamwidth and directivity parameters of the generated dual beams at 1.0 THz here. In case 3, we have $\Omega_{theo.1} = 8.6°$ and $\Omega_{theo.2} = 8.7°$, $D_{theo.1} = 83.8$ and $D_{theo.2} = 82.3$; $\Omega_{cal.1} = 7.6°$ and $\Omega_{cal.2} = 6.9°$, $D_{cal.1} = 94.4$ and $D_{cal.2} = 104.7$; $\Omega_{exp.1} = 9.4°$ and $\Omega_{exp.2} = 7.4°$, $D_{exp.1} = 76.6$ and $D_{exp.2} = 97.9$, for the two LCP THz beams deflected to 0° and 10.8°, respectively. In case 4, these parameters are $\Omega_{theo.1} = 9.3°$ and $\Omega_{theo.2} = 8.7°$, $D_{theo.1} = 77.7$ and $D_{theo.2} = 82.3$; $\Omega_{cal.1} = 7.9°$ and $\Omega_{cal.2} = 8.5°$, $D_{cal.1} = 91.0$ and $D_{cal.2} = 84.9$; $\Omega_{exp.1} = 9.2°$ and $\Omega_{exp.2} = 5.7°$, $D_{exp.1} = 78.5$ and $D_{exp.2} = 125.6$, for the two LCP THz beams deflected to –22° and 10.8°, respectively.

**Single beam steering for slit imaging**

Another relevant application of phased arrays is for imaging.[34] THz imaging plays a crucial role in various applications, including active security screening and location detection. Here, we furthermore showcase the capabilities of our PTPA for slit imaging. As shown by the schematic in Fig. 5a, we positioned a metal slit with 5 mm opening after the PTPA (see Supplementary Note 1). Since the slit only permits THz waves to pass through the opening, we can determine its position by gradually adjusting the deflection angle of the generated THz beam. For simplicity, the schematic focuses solely on the beam steering for LCP THz generation. To effectively image the slit, the angle range of the LCP THz beam must encompass the slit area. According to Eq. (1), we designed 29 single phase gradients to steer the LCP THz beam at 1.0 THz from –29.8° to 29.8°, as detailed in Tab. S1. Figures 5b and c display the 3rd, 15th, and 27th DMD coding patterns and their corresponding phase distributions. In the experiments, these patterns are varied sequentially. Figure 5d shows the measured imaging results of the slit at two different positions. These results are synthesized by extracting the intensity at 1.0 THz and mapping the 29 deflection angles to positions on the slit plane. The gray regions on the left represent areas outside the measurement range of the THz beam with the applied phase gradient. The shaded areas indicate the actual positions of the slit at −12 mm and 8 mm, respectively. The intensity profile clearly reflects the slit positions, although the profiles are wider due to the small size of our PTPA, which causes a broad beamwidth (see Figs. 3f and g) and the diffraction effect of the slit. This setup allows THz signals to be detected beyond the slit area. Increasing the number of elements could improve imaging resolution and reduce such effects.

An important aspect of beam steering is the scanning speed. To assess this metric of performance, we conducted experiments using two DMD coding patterns, indexed by 15th and 27th in Fig. 5b, which support normal and oblique THz emissions. We measured the peak amplitudes with the slit positioned at 0 mm and 18.8 mm, corresponding to the deflection angles of 0.0° and 17.1°, respectively. By holding the delay line at the time-domain peak positions of the THz signals, we observed changes during pattern switching. Figure 5e shows the relationship between the THz peak amplitudes and the switching periods of 1000 ms, 500 ms, 200 ms, and 100 ms. Clear oscillations are noticeable when the switching period reduces to 100 ms. Ideally, the scanning speed of our PTPA is limited only by the switching speed of the DMD coding patterns, which has a maximum binary pattern rate of 4225 Hz. However, our experimental setup's performance is constrained by the laser repetition rate (1.0 kHz), chopper speed (370 Hz), and lock-in integration time (30 ms). Consequently, the full potential of our PTPA is not entirely realized. This limitation could be significantly mitigated by using high repetition rate laser, perfectly continuous lasers for continuous THz generation, on the basis of low threshold THz generation and detection mechanisms. It is also observed that the modulation range measured at 0° is larger because both LCP and RCP THz waves contribute at 0°, whereas only LCP THz waves contribute at 17.1° for pattern 27. Overall, our results effectively demonstrate the potential of our PTPA method for advanced THz imaging applications.

**Programmable THz vortex beam generation**

So far, we have demonstrated the 1D phase control capability of our PTPA. To further showcase its 2D phase control ability, we focus on vortex beam generation, which holds promise for enhancing the spatial channels of wireless communications.[35] Our PTPA enables the generation of THz vortex beam carrying controllable topological charge by designing the pattern of the pump beam, as illustrated in Fig. 6a. A vortex beam's momentum consists of spin angular momentum ($s$) and orbital angular momentum ($l$). Here, $s = +1$ and $-1$ correspond to LCP and RCP, while $l$ denotes the topological charge. Since the square area of phase control is sufficient to generate vortex beam, only half of the fabricated PTPA is utilized. The required interfacial phase distribution should follow:

$$\varphi(x, y) = l\phi, \qquad (4)$$

where $\phi$ represents the azimuth angle of coordinate $(x, y)$. Figures 6b and d illustrates two DMD coding patterns, denoted as case 6 and case 7, for generating vortex beams with $(s, l) = (+1, +1), (-1, -1)$ and $(+1, +2), (-1, -2)$, respectively. Their equivalent phase distributions are depicted in Figs. 6c and e. These coding patterns effectively divide the PTPA into 4 and 8 azimuthal angle ranges, respectively, with each angle range corresponding to one specific phase code.

Figures 6f and h show the calculated the far-field transverse LCP THz radiation intensity and phase distributions at four exemplary frequencies of 0.8, 1.0, 1.2, and 1.4 THz using Rayleigh-Sommerfeld diffraction theory for case 6 and case 7, respectively. These calculations reveal clear donut-shaped intensity profiles and helical phase loops of $2\pi$ and $4\pi$, corresponding to THz vortex beams of $l = 1$ and 2. The deviations from perfect circular shapes in the vortex beams are primarily due to the limited number of azimuthal angle ranges used. Figures 6g and i show the experimental results of the generated LCP THz beams, which exhibit the same vortex beam characteristics with the calculations, demonstrating the 2D phase control ability of our PTPA very well. Additionally, the measured intensity and phase distributions of the generated RCP THz beams are shown in Fig. S5, illustrating THz vortex beams of $l = -1$ and $-2$.

## Discussion

Our results demonstrate the flexibility and programmability of our PTPA up to 0.6-THz bandwidth and 2D operation. Despite limitations in signal-to-noise ratio within our experimental setup, which causes some fluctuations in the measured outcomes, the proof-of-concept demonstrations of our PTPA appear promising for applications. Given that the THz emission cell—or sub-element—is composed of a split-ring resonator (SRR) array, its interaction with adjacent sub-elements is sufficiently minimized to be practically negligible. The nonlinear PB phase underpins the robustness of our phase modulation strategy, allowing us to treat each sub-element as an ideal THz source with a predetermined phase response. The constraint imposed by the sample finite size forces the generated beams to expand upon emission, in turn limiting the achievable directivity. This constraint can be mitigated through the expedient of increasing the aperture size. To validate the efficacy of our approach, the DMD coding patterns implemented in our experiments are derived through simple and direct phase coding methods. The application of more sophisticated phase coding strategies offers promising avenues for further enhancing the performance of our system.

While the current implementation of PTPA uses 2-bit phase coding, the concept can be extended to multiple-bit phase control by increasing the number of sub-elements with varying orientations within each element. The primary limitations are the resolution and size of the DMD, as well as the THz generation efficiency of the sub-elements. These challenges can be addressed by integrating appropriate imaging systems and introducing efficient mechanisms.[36,37] In addition to phase coding, simultaneous amplitude coding is achievable by adjusting either the number of excited sub-elements within each element or the number of active micromirrors within each sub-element. The amplitude can be made proportional to the duty cycle of the local pump beam relative to the sub-elements. Alternatively, exciting multiple sub-elements with different orientations can introduce interference and polarization combinations, thereby offering additional degrees of freedom for more versatile THz applications.

Current THz modulation devices for communications are predominantly based on electric methods, which tend to be bulky. Our PTPA strategy, however, enables the theoretical fusion of optical and THz communications, allowing information transmitted by optical waves to be converted into THz waves. Due to the broadband nature of this method, it supports a significantly larger communication bandwidth and data transfer rate. Moreover, the programmable capability of the PTPA also permits its use as a beamforming device, enabling the delivery of information to one or more targeted end users.

## Conclusion

The development of high-frequency phased arrays is of relevant interest. As we push toward THz frequencies, traditional electronic phased arrays face numerous challenges, but photonics offers a new perspective. In this work, we demonstrated a programmable broadband 2-bit PTPA based on a nonlinear PB phase metasurface. By altering the coding pattern of the pump beam via a DMD, our PTPA achieves real-time 2D wavefront control of THz waves during their generation. Our method supports key functionalities such as dynamic THz beamforming, precise beam steering for imaging, and vortex beam generation. A significant advantage of our PTPA is its integration of wavefront control directly into the generation process, which eliminates the need for additional THz insertion and mitigates bandwidth losses. Furthermore, the principles demonstrated in this study can be applied to other nonlinear THz generation platforms, as well as to linear metasurface platforms controlled by optical pumps. This adaptability enables the realization of similar programmable functionalities across various systems. Our research thus represents a transformative shift in THz technology, providing unprecedented levels of flexibility, adaptability, and functionality for a wide array of THz applications.

## Methods

### Fabrication

The nonlinear metasurface for PTPA were fabricated through a three-step electron beam lithography process, involving electron beam lithography, electron beam evaporation, and lift-off. Initially, the PTPA patterns were created on a 0.7 mm-thick fused quartz substrate coated with an 8 nm-thick ITO layer using a JEOL 9300FS 100 kV EBL tool. The thin ITO layer serves dual purposes of preventing local charge accumulation during lithography and enhancing THz generation. Subsequently, 3 nm-thick Ti and 40 nm-thick Au films were successively deposited via e-beam evaporation. Finally, lift-off and cleaning processes were executed to obtain the PTPA.

**Acknowledgements**

This work is supported by the National Natural Science Foundation of China (Grant Nos. 62075158 (X.Z.), 62025504 (J.H.), 61935015 (J.H.), 62405215 (X.C.) and 62135008 (Q.X.)); China Postdoctoral Science Foundation (2024M752359 (X.C.)); Postdoctoral Fellowship Program of CPSF (GZC20241200 (X.C.)); Yunnan Expert Workstation (Grant No. 202205AF150008 (J.H.)). A.A. was supported by the Simons Foundation.

**Supplementary Materials**

Supplementary materials for this article is available at http://xxxx.

**Conflict of interest**

The authors declare no competing interests.

**Contributions**

X.Z. conceived the idea. X.Z., L.N. and X.F. contributed to project conceptualization, methodology, and validation; L.N. and X.F. performed the measurements; L.N. fabricated the samples; L.N. and X.Z. wrote the manuscript with suggestions from Y. L., Q.W., Q.X., X.C, J.M., H.Q., W.E.I.S., S.Z., A.A, W.Z and J.H. X. Zhang and J. Han supervised the project. All authors discussed the results and prepared the manuscript.

# Figure captions

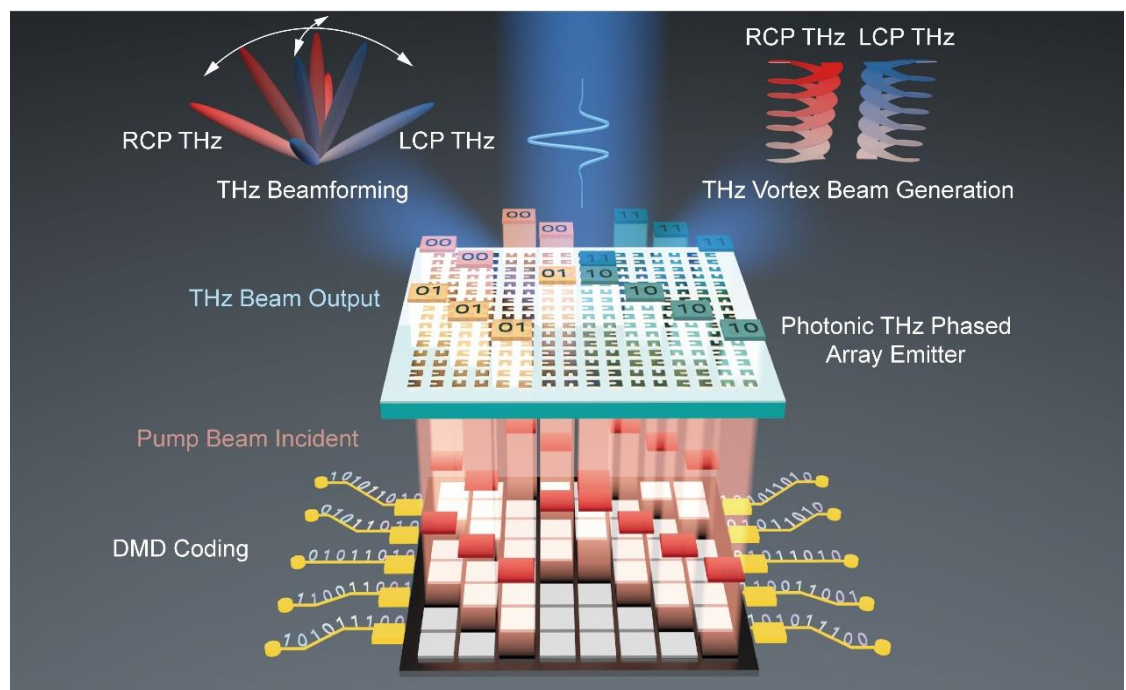

**Fig. 1 Concept of PTPA.** The PTPA is composed of SRRs with four different orientations, corresponding to 2-bit phase coding to the generated THz waves represented by "00", "01", "10" and "11". By inputting desired "0" and "1" coding patterns to the DMD, the angle state of each micromirror can be controlled. The red beam represents the coded near-infrared femtosecond pump beam pattern by the DMD, where only the beam parts reflected by the micromirrors controlled in the schematic black regions can be directed to the PTPA. By matching this beam pattern with SRRs of selected orientations, arbitrary 2-bit phase distribution of the generated THz waves can be generated, thus allowing programmable control over the propagation of the THz waves. The top insets show three proof-of-concept demonstrations of the PTPA, including beamforming and vortex beam generation.

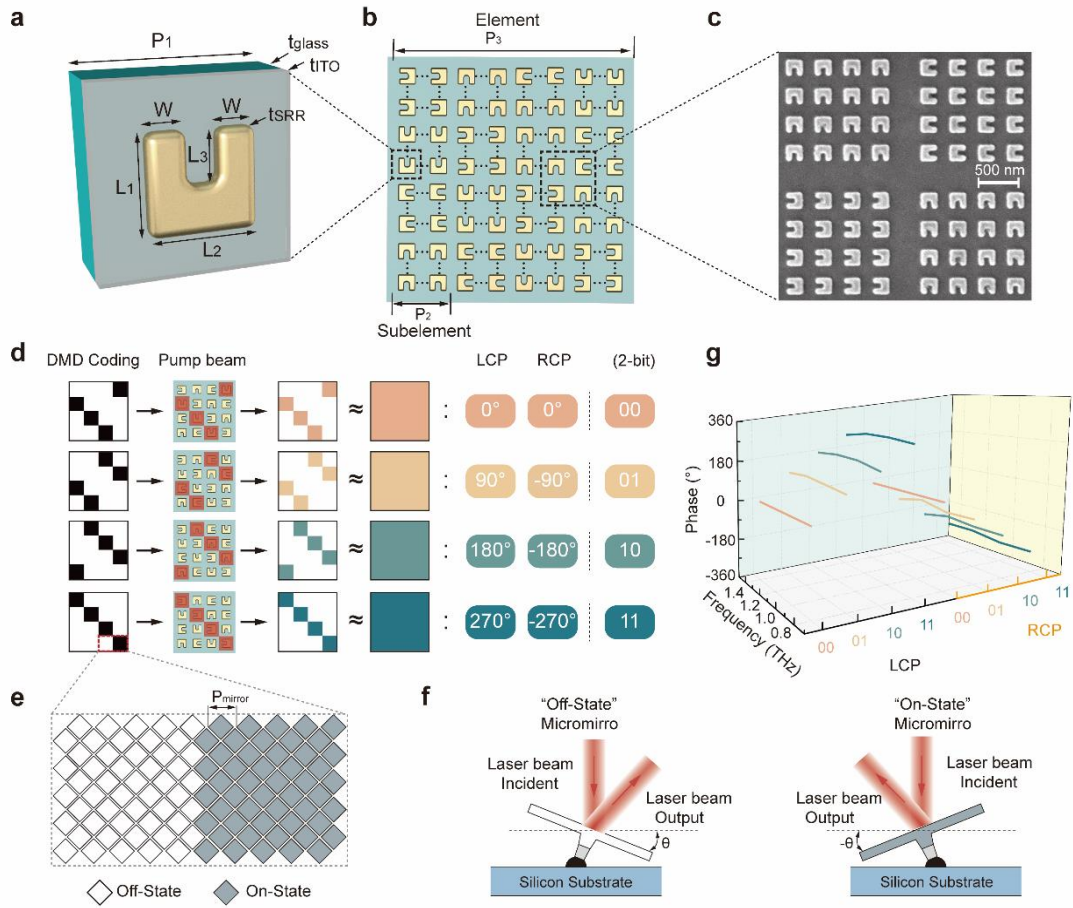

**Fig. 2 Design of the PTPA element. a** Schematic of the SRR unit cell, which has dimensions of $L_1$ = 212 nm, $L_2$ = 220 nm, $L_3$ = 110 nm, $W$ = 79 nm, $t_{SRR}$ = 23 nm, $t_{ITO}$ = 8 nm, $t_{glass}$ = 0.7 mm and $P_1$ = 382 nm, respectively. **b** Schematic of the PTPA element, which is composed of 4 × 4 subelements with four different orientations of 0°, 90°, 180° and 270°. Each orientation corresponds to four subelements placed in a staggered arrangement. The periods of each subelement and element is $P_2$ = 50 μm and $P_3$ = 200 μm, respectively. **c** Partial scanning electron microscope image of the fabricated sample. **d** DMD coding patterns to the local pump beam illuminating on one element, which selectively excited the subelements to achieve 2-bit nonlinear PB phase control over the generated LCP and RCP THz waves. **e** Enlarged view of the micromirrors' states of the white and black regions in **d**, corresponding to off-state and on-state, respectively. **f** The angles of micromirrors working in off-state and on-state. **g** Measured normalized broadband phase responses of LCP and RCP THz waves under different DMD coding patterns in **d**.

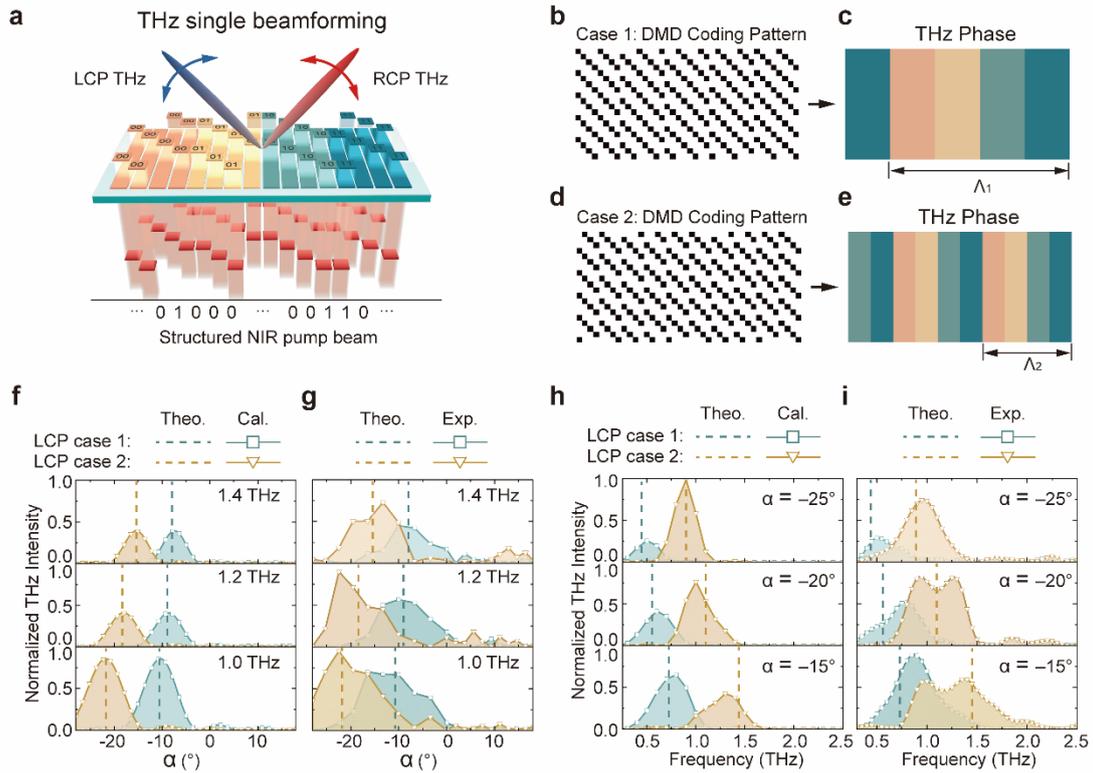

**Fig. 3 Programmable THz single beamforming. a** Schematic of the PTPA functionality on single beamforming. **b,d** DMD coding patterns to achieve two different linear phase gradients along the *x* directions, which are denoted as case 1 and case 2. **c,e** Equivalent phase distributions induced by the DMD coding patterns in **b,d**. **f,g** Calculated and measured angle-resolved intensity distributions of the generated LCP THz waves at 1.0, 1.2 and 1.4 THz for case 1 and case 2, respectively. **h,i** Calculated and measured intensity spectra of the generated LCP THz waves at −25°, −20° and −15° for case 1 and case 2, respectively. The dashed lines are theoretical deflection angle and frequency results using generalized Snell's law.

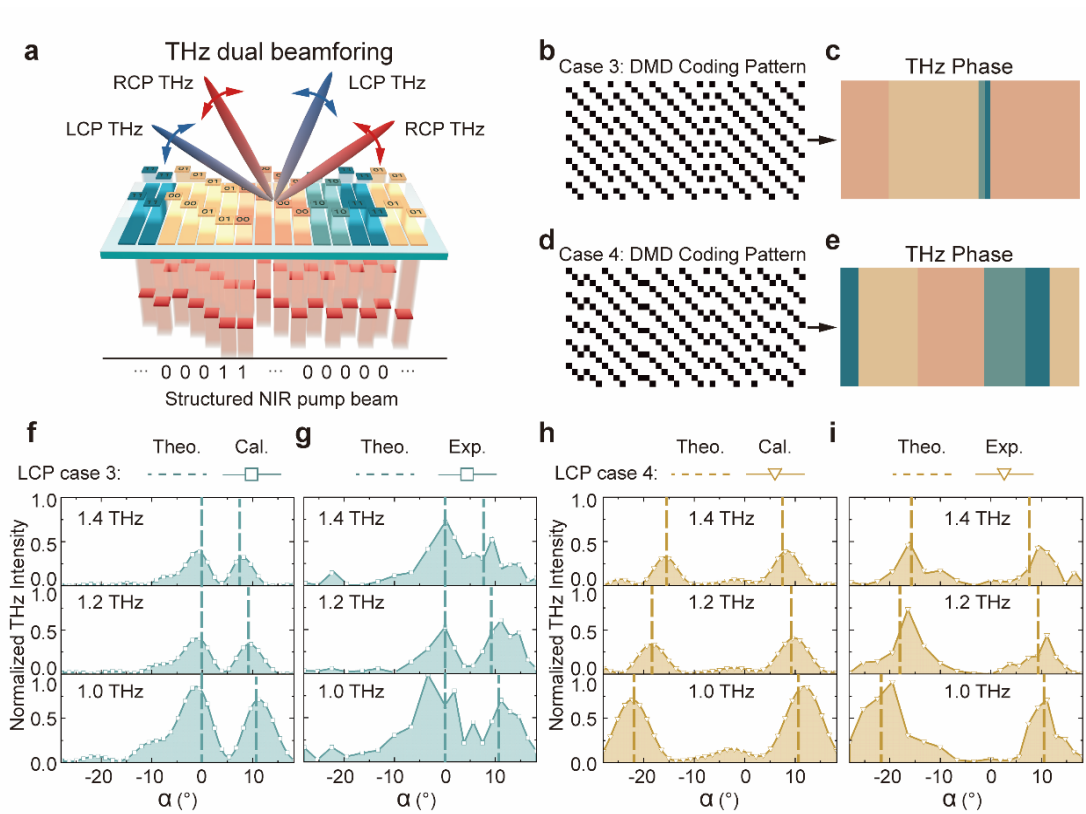

**Fig. 4 Programmable THz dual beamforming. a** Schematic of the PTPA functionality on dual beamforming. **b,d** DMD coding patterns to achieve two different phase distributions individually merged by two different phase gradients along the *x* directions, which are denoted as case 3 and case 4, respectively. **c,e** Equivalent phase distributions induced by the DMD coding patterns in **b,d**. **f,h** and **g,i** Calculated and measured angle-resolved intensity distributions of the generated LCP THz waves for case 3 and case 4, respectively.

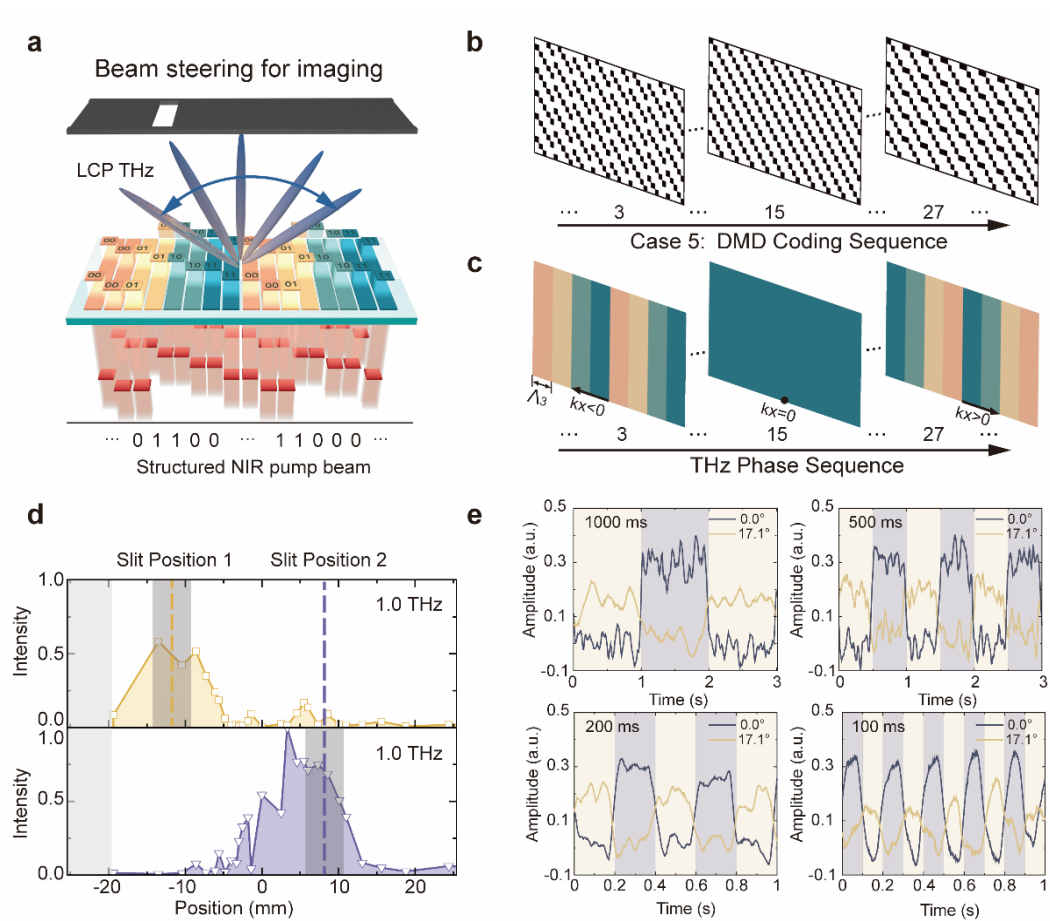

**Fig. 5 Single beam steering for slit imaging. a** Schematic of the imaging mechanism. **b** DMD coding patterns to achieve gradual THz beam steering along the *x* direction. **c** Equivalent phase distributions induced by the DMD coding patterns in **b**. **d** Measurements intensity profiles of the generated LCP THz waves at 1.0 THz when the slit are placed at two different positions. The horizontal axis is obtained by transforming the deflection angles at 1.0 THz to positions at the slit plane based on the 29 phase gradients in Tab. S1. **e** Measured variations of the THz time-domain peak amplitudes at 0.0° and 17.1° under different switching periods of the DMD coding patterns (15 and 27), respectively.

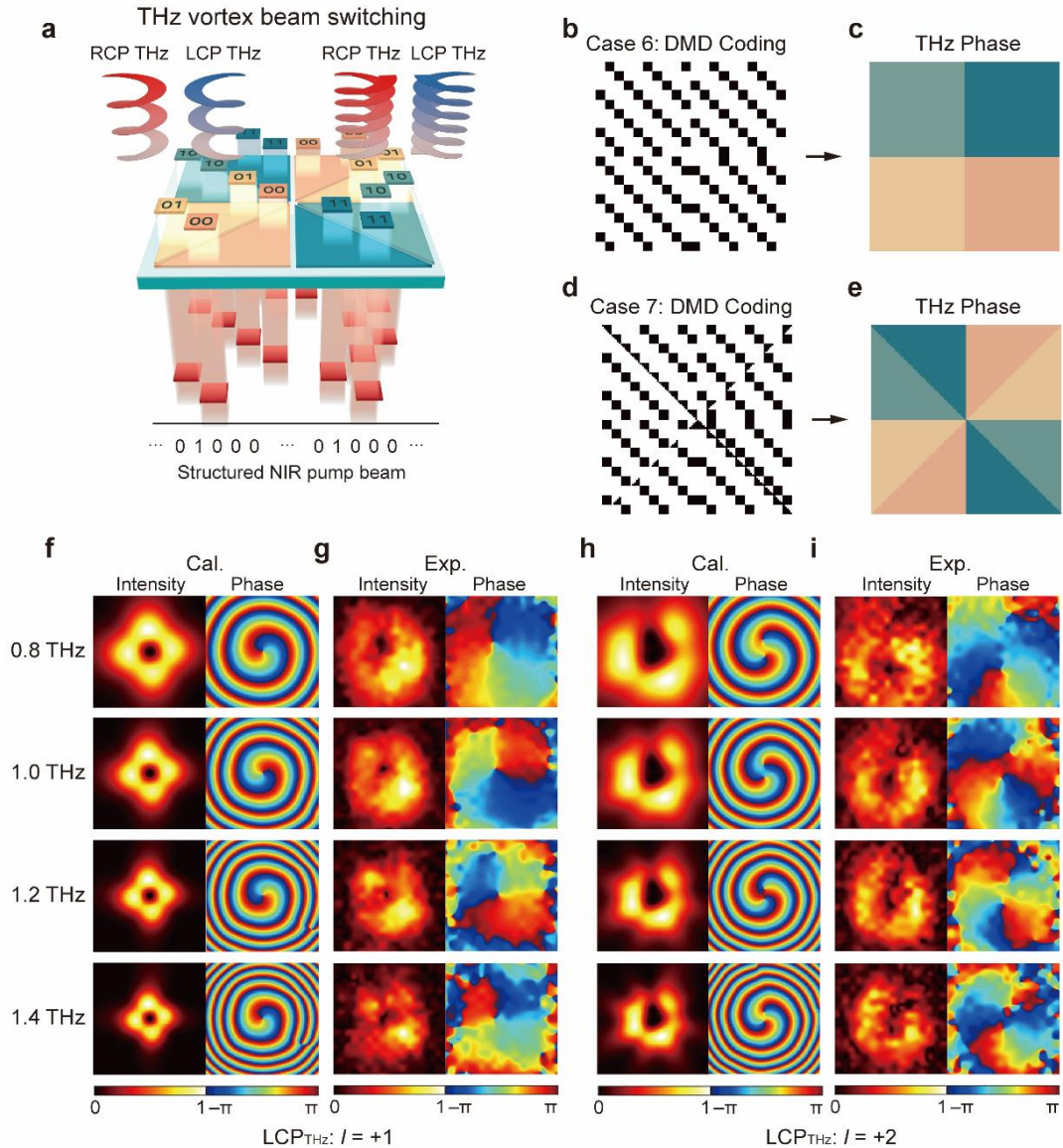

**Fig. 6 Programmable THz vortex beam generation. a** Schematic of the PTPA functionality on THz vortex beam generation. **b,d** DMD coding patterns to achieve helical phase distributions of topological charges $|l| = 1$ and $|l| = 2$, which are denoted as case 6 and case 7, respectively. **c,e** Equivalent phase distributions induced by the DMD coding patterns in **b,d**. **f,h** and **g,i** Calculated and measured transverse intensity and phase distributions of the generated LCP THz beams at 0.8, 1.0, 1.2, and 1.4 THz for case 6 and case 7, respectively.